\documentclass[fleqn,usenatbib]{mnras}
\usepackage{newtxtext,newtxmath}
\usepackage[T1]{fontenc}
\usepackage{ae,aecompl}
\usepackage{relsize}

\usepackage{graphicx}	
\usepackage{amsmath}	
\usepackage{upgreek}
\usepackage{bigints}
\usepackage{bm}

\graphicspath{{./}{figures/}}

\title[TDE stream self-intersection]{An Analytic, Fully Relativistic Framework for Tidal Disruption Event Streams in Schwarzschild Geometry }

\author[A. J. Dittmann]{
Alexander J. Dittmann$^{1}$\thanks{E-mail: \href{mailto:dittman@astro.umd.edu}{dittmann@astro.umd.edu}}
\\
$^{1}${Department of Astronomy and Joint Space-Science Institute, University of Maryland, College Park, MD 20742-2421}\\
}

\date{\today}

\pubyear{2021}

\begin{document}
\label{firstpage}
\pagerange{\pageref{firstpage}--\pageref{lastpage}}
\maketitle

\begin{abstract}
We present an analytic and fully relativistic framework for studying the self-intersection of tidal disruption event (TDE) streams, restricting ourselves to the Schwarzschild spacetime. By taking advantage of the closed-form solution to the geodesic equations in the Schwarzschild metric, we calculate properties of the self-intersection without numerically evaluating the geodesic equations or making any post-Newtonian approximations. Our analytic treatment also facilitates geometric definitions of the orbital semi-major axis and eccentricity, as opposed to Newtonian formulas which lead to unphysical results for highly-relativistic orbits. Combined with assumptions about energy dissipation during the self-intersection shock, our framework enables the calculation of quantities such as the fraction of material unbound during the self-intersection shock, and the characteristic semi-major axes and eccentricities of the material which remains in orbit after the collision. As an example, we calculate grids of post-intersection properties in stellar and supermassive black hole (SMBH) masses for disruptions of main sequence stars, identifying regions where no material is ejected during self intersection (e.g. SMBH mass $\lesssim 5\times10^6\, {\rm M_\odot}$ for $1\,{\rm M_\odot}$ stars disrupted at the tidal radius), potentially explaining the TDEs observed by SGR/eROSITA which are visible in X-rays but not optical wavelengths. We also identify parameters for which the post-intersection accretion flow has low eccentricity ($e\lesssim0.6$),
and find that the luminosity generated by self-intersection shocks only agrees with observed trends in the relationship between light curve decay timescales and peak luminosities over a narrow range of SMBH masses.
\end{abstract}



\section{Introduction}
A tidal disruption event (TDE) occurs when the halcyon days of an unlucky star are brought to an abrupt end by a strong tidal encounter with a supermassive black hole (SMBH). Such encounters can illuminate previously quiescent galactic nuclei \citep[e.g.][]{1979SvAL....5...16L,1988Natur.333..523R}, and provide a method to study lower-mass SMBHs than can be probed using dynamical mass measurements \citep[e.g.][]{2000ApJ...539L...9F,2001MNRAS.320L..30M,2002ApJ...574..740T,2009ApJ...698..198G,2019ApJ...872..151M}, although some lower-mass SMBHs can be studied in Seyfert galaxies \citep[e.g.][]{2011ApJ...739...28X}. Partial disruptions of evolved stars may provide a quasi-steady source of material to SMBHs, seeding comparable material to stellar winds and powering low-luminosity flares \citep{2013ApJ...777..133M}. Tidal disruptions may also occur in active galactic nuclei \citep[e.g.][]{2019ApJ...881..113C,2020ApJ...903...17C,2021arXiv211003741M}, as hinted at by some observations \citep[e.g.][]{2017ApJ...843..106B,2018ApJ...859....8L,2020ApJ...894...93L}. 

Although the first candidate TDEs were detected as soft X-ray transients in otherwise-quiescent galactic nuclei by the ROSAT All-Sky Survey \citep{2002AJ....124.1308D}, the identification of TDE candidates has become a hallmark of large optical surveys such as Pan-STARRS \citep{2012Natur.485..217G,2014ApJ...780...44C,2019ApJ...880..120H,2019MNRAS.488.1878N}, the intermediate Palomar Transient Factory \citep{2017ApJ...842...29H,2017ApJ...844...46B,2019ApJ...873...92B}, the All-Sky Automated Survey for Supernovae (ASAS-SN) \citep{2014MNRAS.445.3263H,2016MNRAS.463.3813H,2016MNRAS.455.2918H,2019ApJ...883..111H,2019MNRAS.488.4816W}, and the Zwicky Transient Facility (ZTF) \citep{2019ApJ...872..198V,2021ApJ...908....4V}. One intriguing characteristic of many of these events is that the inferred blackbody radius of the ultraviolet and optical emission tends to be factors of $\sim10-100$ times larger than the size of the discs expected to form after the circularisation of TDE debris, suggesting the presence of winds \citep[e.g.][]{2015ApJ...805...83M,2018ApJ...859L..20D} or unbound material from the self-intersection of disrupted debris streams \citep[e.g.][]{2015ApJ...806..164P,2016ApJ...830..125J,2020MNRAS.492..686L}. Out of these TDEs selected via optical variability, only a minority have been accompanied by detectable X-ray emission \citep[e.g.][]{2021ApJ...908....4V}, which may be due to obscuration and reprocessing of X-rays emitted as gas accretes onto the SMBH by outflowing material \citep[e.g.][]{2018ApJ...859L..20D,2020MNRAS.492..686L}. On the other hand, the SGR/eROSITA all-sky X-ray survey has recently identified a population of TDEs without any detectable optical emission \citep{2021MNRAS.508.3820S}. These observations, among others, motivate a more comprehensive understanding of the TDE stream self-intersection process.

After stars are ripped apart by the tidal forces of the SMBH, stellar debris returns at a rate determined by how bound a given fluid parcel is to the SMBH \citep[$ \propto t^{-5/3} $,][]{1989IAUS..136..543P}. For a canonical parabolic orbit, roughly half of the stellar material remains bound to the SMBH while the other half escapes the system. The most bound material, that which was closest to the SMBH during pericentre passage, returns first having the lowest energy, smallest semi-major axis, and thus shortest orbital period. 
As the disrupted, bound material again passes by the SMBH, it experiences general relativistic apsidal precession, causing material outgoing from the SMBH to intersect with incoming material, leading to disc formation \citep[see, e.g.][]{2015ApJ...804...85S,2016MNRAS.455.2253B,2016MNRAS.458.4250S}. 
The self-intersection process and subsequent accretion disc formation could be responsible for generating the observed TDE optical emission \citep[e.g.][]{2015ApJ...806..164P,2016ApJ...830..125J,2017MNRAS.467.1426S}, determining the size of the X-ray emitting accretion disc \citep[e.g.][]{2015ApJ...812L..39D}, and launching obscuring outflows \citep[e.g.][]{2020MNRAS.492..686L}.

Previous studies of the stream self-intersection process have relied on post-Newtonian approximations \citep[e.g.][]{2015ApJ...812L..39D,PhysRevD.104.103019,2021arXiv211109173W}, or resorted to numerical evaluation of the geodesic equations to determine the stream properties at self-intersection while retaining a fully relativistic treatment \citep[e.g.][]{2020MNRAS.492..686L,2021arXiv211203918B}. We detail in Section \ref{sec:geodesics} a treatment of the stream self-intersection process which is both fully relativistic and completely analytic, to determine quantities such as the radius of stream self-intersection, under the assumption of a non-spinning SMBH. We apply this framework to different phenomenological models of the physics of the self-intersection shock, investigating completely inelastic collisions in Section \ref{sec:dissipation} and more general collisions in Section \ref{sec:shocks}. We comment on the implications of our model for accretion flow dynamics and in light of recent observations, and discuss limitations of our method in Section \ref{sec:discuss}, and conclude in Section \ref{sec:conclusions}. Unless otherwise specified, in this work we use geometric units where $G=1,~c=1$, and express all lengths in units of the gravitational radius $r_g=M_\bullet$, where $M_\bullet$ is the SMBH mass.

\section{TDE streams as Schwarzschild Geodesics}\label{sec:geodesics}
We begin by relating astrophysical quantities such as the masses of SMBHs and the stars they disrupt, stellar radii, and orbit pericentres, to conserved quantities in Schwarzschild spacetime. We consider a star of mass $M_*$ and radius $R_*$. The classical radius at which tidal forces from the SMBH may unbind the star can be defined as \citep{hill1878researches}
\begin{equation}
r_T = R_*\left(\frac{M_\bullet}{M_*}\right)^{1/3},
\end{equation}
although relativistic effects, realistic stellar structure, and stellar rotation can lead to pericentre distances smaller than $r_T$ being required for full tidal disruptions \citep[e.g.][]{2019ApJ...882L..25L,2020ApJ...904...98R,2020ApJ...904..100R}. We express the pericentre radius of an initial stellar orbit as $r_p=r_T/\beta$, where $\beta$ is a free impact parameter. We specialise to stars on initially-parabolic orbits in this work for the sake of brevity. After a star is disrupted, its debris acquires specific orbital energies ranging from $\mathcal{E}\sim -\eta \mathcal{E}_T$ to $\mathcal{E}\sim \eta\mathcal{E}_T$, where $\eta$ is a constant of order unity and $\mathcal{E}_T$ is the Newtonian tidal energy 
\begin{equation}
\mathcal{E}_T=\frac{R_*M_\bullet}{r_T^2}.
\end{equation}
Throughout this work we set $\eta=1$ for simplicity. 

Hereafter we will consider the orbital behaviour of the most-bound debris with specific orbital orbital energy $-\mathcal{E}_T$, the leading edge of the stream. As the width of the stream is thought to be much smaller than the pericentre radius \citep[e.g.][]{1994ApJ...422..508K,2016MNRAS.455.3612C}, a given portion of the stream moves along a geodesic until colliding with another segment of the stream. Each geodesic is 
characterised by two constants of motion: the specific energy
$E=1+\mathcal{E}=1-\mathcal{E}_T$
and specific angular momentum $h$.

The geodesic equations are then, expressing derivatives of a quantity $q$ with respect to the proper time of a particle on that geodesic $\tau$ as $\dot{q}$,
\begin{align}
\dot{t} = \frac{E}{1-r_s/r}, \label{eq:tdot}\\
\dot{\phi} = \frac{h}{r^2}, \label{eq:phidot} \\
\dot{r}^2 = E^2 - \left(1-\frac{r_s}{r} \right)\left(1+\frac{h^2}{r^2} \right), \label{eq:rdot2}\\
\ddot{r} = -\frac{r_g}{r^2} + \frac{h^2}{r^3}\left(1-\frac{3r_g}{r} \right), \label{eq:rddot}
\end{align}
where $r$ is the circumferential radius, $\phi$ is a longitude, $t$ is the time measured by a stationary clock at infinity, and $r_s=2r_g$.
We note that at pericentre, $\dot{r}=0$, in which case Equation (\ref{eq:rdot2}) yields an expression for the specific angular momentum of a geodesic with a given energy and pericentre distance 
\begin{equation}
h = r_p\sqrt{\frac{E^2}{1-r_s/r_p}-1}.
\end{equation}\label{eq:h}
When the specific angular momentum is nonzero, Equations (\ref{eq:phidot}) and (\ref{eq:rdot2}) can be rewritten as
\begin{equation}
\left(\frac{dr}{d\phi}\right)^2 = \frac{r^4E^2}{h^2} - \left(1-\frac{r_s}{r}\right)\left(\frac{r^4}{h^2}+r^2\right),
\end{equation}
and then in terms of the inverse radius $u\equiv1/r$ as 
\begin{equation}\label{eq:dueq}
\left(\frac{du}{d\phi} \right)^2 = \frac{E^2}{h^2} - (1-r_su)\left(\frac{1}{h^2}+u^2\right).
\end{equation}
The solution to equation (\ref{eq:dueq}) is, in terms of its roots labelled from smallest to largest as $u_1,u_2,u_3$, and the $\rm sn$ (\textit{sinus amplitudinis}) 
Jacobi elliptic function \citep{1930JaJAG...8...67H}, 
\begin{equation}\label{eq:schworbit}
u(\phi) = u_1 + (u_2-u_1) \rm{sn^2}\left(\frac{1}{2}\phi\sqrt{r_s(u_3-u_1)},k\right),
\end{equation}
where we have chosen the arbitrary constant of integration such that $\phi=0$ corresponds to the apocentre of the geodesic, and the elliptic modulus of $\rm sn$ is $k=\sqrt{(u_2-u_1)/(u_3-u_1)}$. We provide an abbreviated derivation of Equation (\ref{eq:schworbit}) as well as closed-form expressions for the roots of equation (\ref{eq:dueq}) in Appendix \ref{app:deriv}. 

In the interpretation of Equation (\ref{eq:schworbit}), it is useful to note that when all roots of Equation (\ref{eq:schworbit}) are real, $u_1$ is the inverse of the apocentre distance, $u_2$ is the inverse of the pericentre distance, and in the Newtonian limit where $r \gg r_s$, $u_3 \approx 1/r_s \gg u_1, u_2$ such that equation (\ref{eq:schworbit}) reduces to the equation for a standard Keplerian orbit. 

\begin{figure}
\includegraphics[width=\columnwidth]{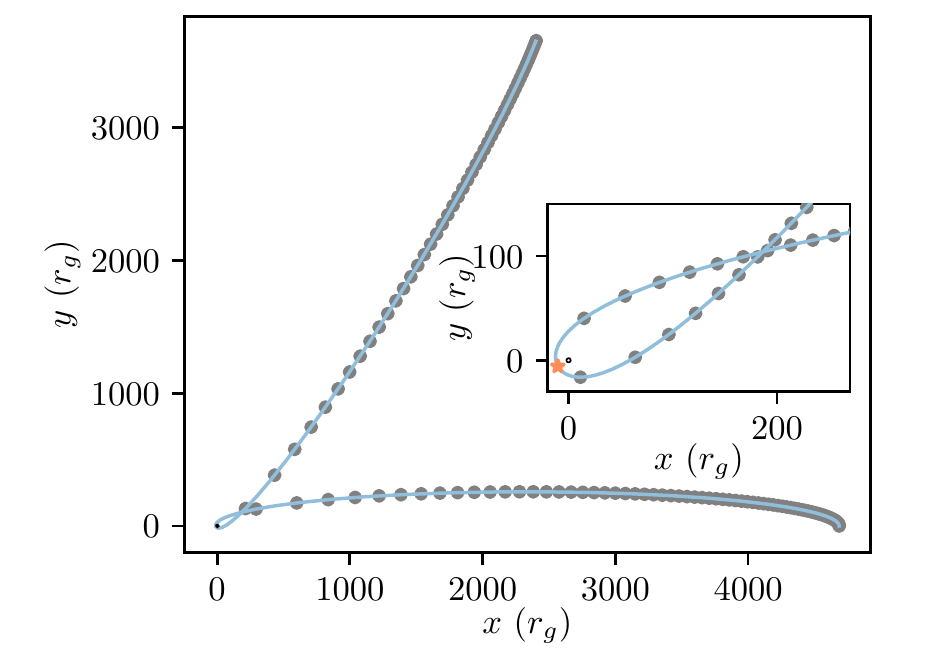}
\caption{A portion of the geodesic (in blue) traced by the most-bound debris of a solar-mass star on a $\beta=4$ orbit disrupted by a $10^6\,M_\odot$ SMBH. The Schwarzschild radius is marked by a black circle. In the inset, the pericentre of the orbit is marked by an orange star symbol. Note that the geodesic intersects itself at $\phi_p\pm\pi$, where $\phi_p$ is the argument of periapsis. The grey points are sampled from a direct integration of the geodesic equations using a 6th-order time-symmetric Hermite scheme \citep{2008NewA...13..498N,2020MNRAS.496.1217D}.}
\label{fig:orbit}
\end{figure}

\begin{figure}
\includegraphics[width=\columnwidth]{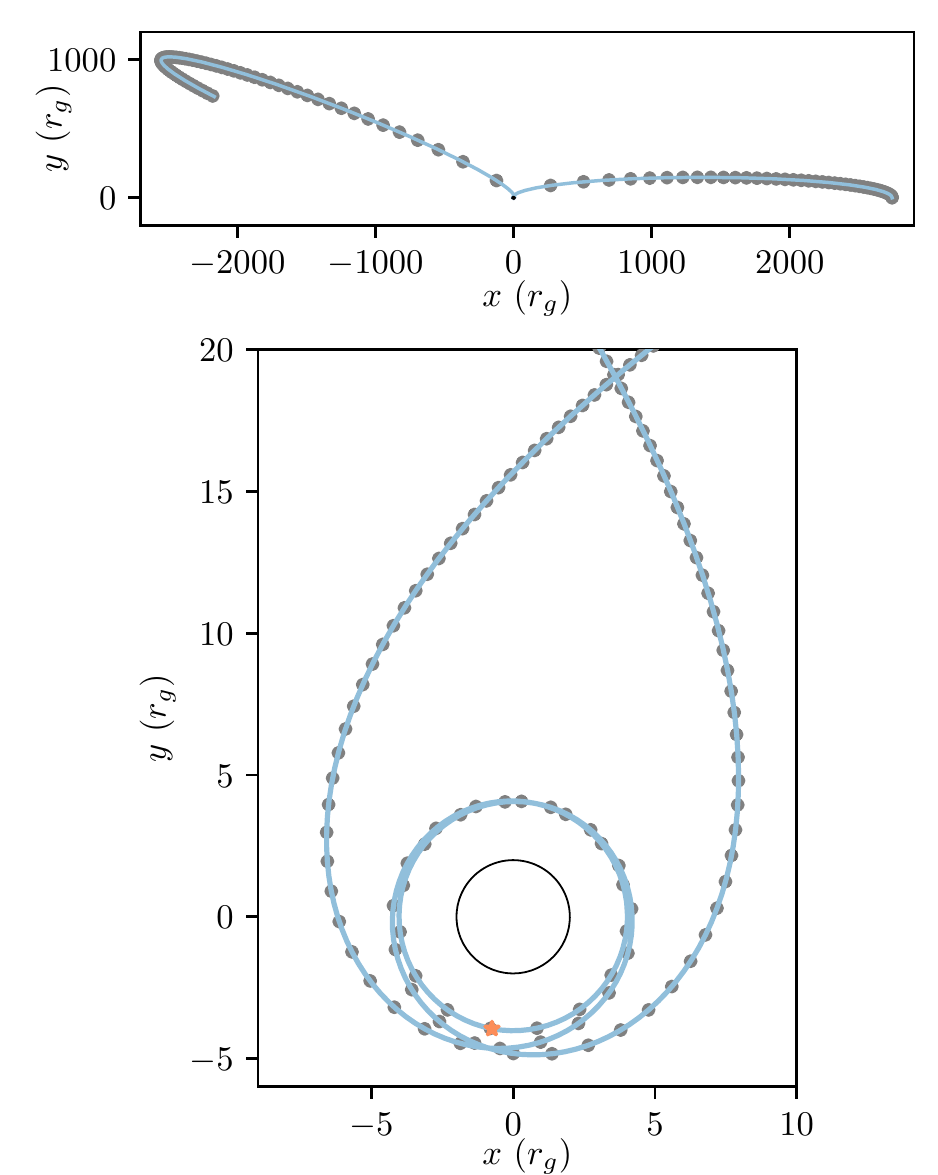}
\caption{Geodesics (in blue) traced by the most-bound debris of a solar mass star on a $\beta=4$ orbit disrupted by a $5\times10^6\,M_\odot$ SMBH. The bottom panel focuses on the orbit pericentre, marked by an orange star symbol, illustrating the three self-intersections per orbit, the first of which occurs nearest to periapse, each occurring $\pi$ radians in $\phi$ from the others. For this geodesic, $u_2$ and $u_3$ are both close to $1/(2r_s)$, so $\mathcal{K}$ and accordingly $\Delta\phi$ become large compared to the values for typical more typical geodesics. The grey points are sampled from a direct integration of the geodesic equations using a 6th-order time-symmetric Hermite scheme \citep{2008NewA...13..498N,2020MNRAS.496.1217D}.}
\label{fig:orbit2}
\end{figure}

The quarter-period of $\rm sn$ and thus the half-period of an orbit described by Equation \ref{eq:schworbit} is the complete elliptic integral of the first kind $\mathcal{K}(k)$. Thus, pericentre occurs at $\phi_p=2\mathcal{K}/\sqrt{r_s(u_3-u_1)}$ and the change in $\phi$ over the course of one oscillation in $r$ or $u$ is $\Delta\phi = 4\mathcal{K}/\sqrt{r_s(u_3-u_1)}$, and $\Delta \phi - 2\pi$ is the apsidal precession over the course of an orbit. Geometrically, it is clear that the stream must self-intersect at 
\begin{equation}
r_i=r(\phi_p \pm \pi),
\end{equation}
as illustrated in Figures \ref{fig:orbit} and \ref{fig:orbit2}. Note that for orbits with $u_1\ll u_2,u_3$ and $u_2\sim u_3$, $k$ approaches $1$ and $\mathcal{K}$ diverges logarithmically, leading to multiple self-intersections each pericentre passage as illustrated in Figure \ref{fig:orbit2}.

Before examining characteristics of the stream intersections for grids of parameters, it is useful to examine the properties of Equation (\ref{eq:schworbit}) and its roots. For oscillations of Equation (\ref{eq:schworbit}) to be well-defined, $u_1,u_2,u_3$ must be distinct. From Equation \ref{eq:dueq}, we observe that $u_1+u_2+u_3=1/r_s$ and $u_1u_2u_3=(1-E^2)/(h^2r_s)$ \citep{girard1884invention}. From the latter, along with Descartes' rule of signs, it is clear that for \emph{bound} orbits with three distinct real roots, all roots are positive. Then, for the roots to be distinct, the pericentre distance must be greater than $4r_g$, so that $u_3$ can be greater than $u_2$ while the roots still sum to $1/r_s$. For this reason, we only consider orbits such that $r_p>4r_g$. Similarly, the smallest possible circular geodesic has $u_1=u_2=u_3=1/6r_g$.

Once the self-intersection radius is known, it is straightforward to calculate velocities at the self-intersection point from Equations (\ref{eq:phidot}) and (\ref{eq:rdot2}). Following \citet{2020MNRAS.492..686L}, we calculate velocities in the frame of a stationary observer at $r_i$, define $\mu_i\equiv1-r_s/r_i$, and denote quantities $q$ measured in this frame by $\widetilde{q}$. The local differential time is then $d\widetilde{t}=E\mu_i^{-1/2}d\tau$, and the local differential length in the radial direction is $d\widetilde{r}=\mu_i^{-1/2}dr$. The radial and azimuthal velocities at the intersection point are then given by
\begin{equation}\label{eq:vsquigr}
\widetilde{v}_r = \frac{d\widetilde{r}}{d\widetilde{t}}=\frac{\dot{r}(r_i)}{E},
\end{equation}
\begin{equation}\label{eq:vsquigp}
\widetilde{v}_\phi = r_i\frac{d\widetilde{\phi}}{d\widetilde{t}}=\frac{r_i\mu_i^{1/2}}{E}\dot{\phi}(r_i),
\end{equation}
and the intersection half angle $\widetilde{\theta}_i$ is given by
\begin{equation}
\tan{\widetilde{\theta}_i}=\frac{\widetilde{v}_r}{\widetilde{v}_\phi}=\frac{\dot{r}(r_i)}{\nu_i^{1/2}r_i\dot{\phi}(r_i)}.
\end{equation}

In order to make comparisons to previous works, we review the post-Newtonian approximations of $\widetilde{\theta}_i$, $\Delta\phi-2\pi$, and $r_i$ presented in \citet{2015ApJ...812L..39D}. In the limit where $u_3 \approx 1/r_s \gg u_1, u_2$ and $k^2\approx r_s(u_2-u_1)\ll 1$,  to lowest order $\mathcal{K}\approx(1+k^2/4)\pi/2$, which leads to $\Delta\phi-2\pi\approx3\pi r_s(u_1+u_2)/2$. Recalling that $u_1$ and $u_2$ are the inverse orbital apocentre and pericentre and $u_1+u_2$ can be written in terms of an orbital semi-major axis $a$ and eccentricity $e$ as $u_1+u_2=2/(a(1-e^2))$, the leading-order post-Newtonian approximation for the apsidal precession is 
\begin{equation}
\Delta\phi-2\pi \approx \frac{3\pi r_s}{a(1-e^2)}.
\end{equation}
In the approximation that this precession occurs instantaneously at pericentre, \citet{2015ApJ...812L..39D} find geometrically that
\begin{equation}
r_i \approx \frac{(1+e)r_p}{1-e\cos{(\Delta\phi/2-\pi)}}
\end{equation}
and
\begin{equation}
2\widetilde{\theta}_i\approx \cos^{-1}\left(\frac{1-2e\cos{(\Delta\phi/2-\pi)}+e^2\cos{(\Delta\phi-2\pi)}}{1-2e\cos{(\Delta\phi/2-\pi)}+e^2}\right).
\end{equation}

\begin{figure}
\includegraphics[width=\columnwidth]{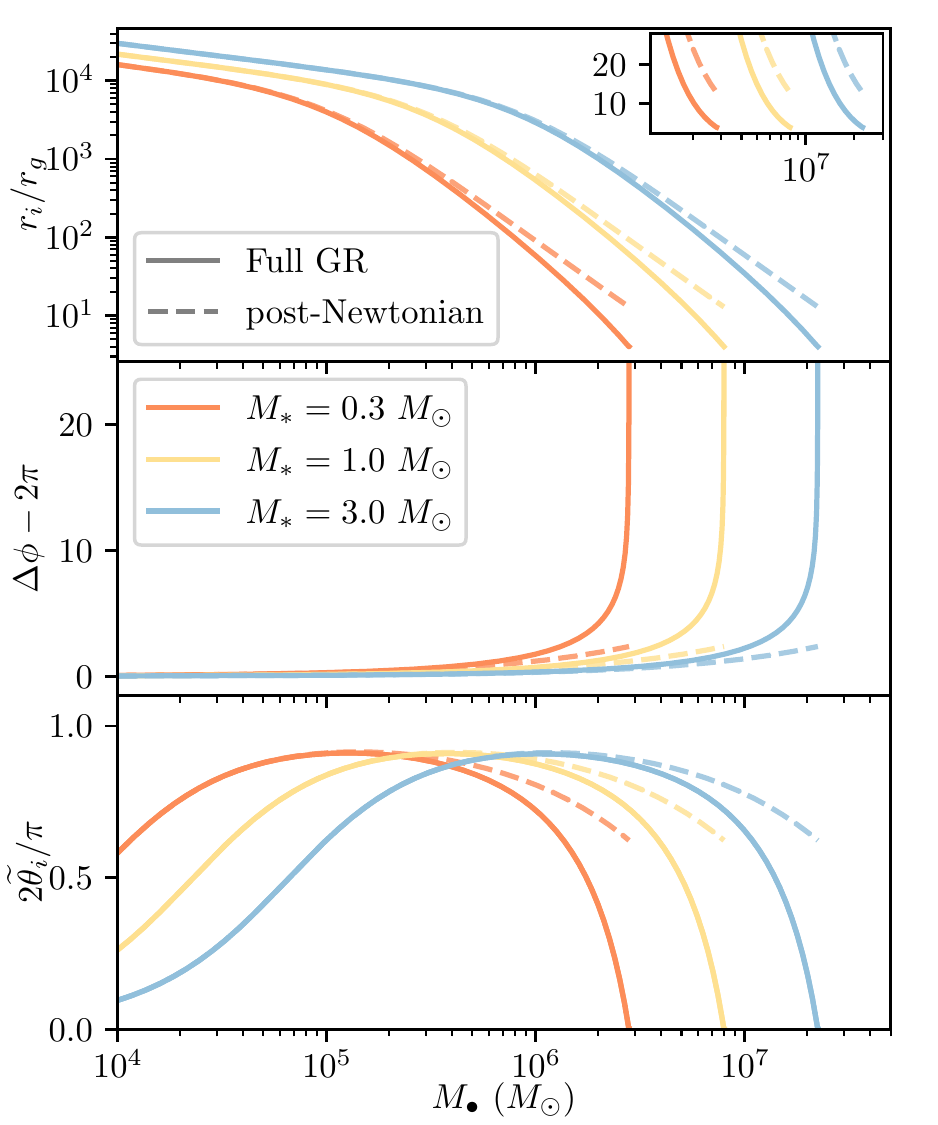}
\caption{The self-intersection radius (top panel), apsidal precession (middle panel), and angle of stream intersection (bottom panel) in full general relativity (solid lines) and in the 1st-order post-Newtonian approximation used in \citet{2015ApJ...812L..39D} (dashed lines) for the geodesics traced by the most bound debris from a disrupted stars of masses $M_*/M_\odot=\{0.3, 1.0, 3.0\}$ plotted in orange, yellow, and blue respectively with penetration factors $\beta=3$. Radii for each star are the intermediate age main sequence values from the MIST v1.2 evolutionary tracks \citep{2016ApJS..222....8D,2016ApJ...823..102C}.}
\label{fig:intersection}
\end{figure}

We compare fully relativistic calculations of the intersection radius, apsidal precession, and angle of stream intersection to the corresponding post-Newtonian approximations in Figure \ref{fig:intersection} for SMBH masses from $10^4~M_\odot$ up the the SMBH mass at which $r_p\approx4r_g$. We consider the disruptions of $0.3$, $1.0$, and $3.0~M_\odot$ stars on $\beta=3$ orbits, using the intermediate age main sequence\footnote{Intermediate age main sequence is defined as the point where the core hydrogen mass fraction drops to $X_c=0.3$ \citep{2016ApJS..222....8D}.} radii from MESA Isochrones and Stellar Tracks (MIST) v1.2 \citep{2011ApJS..192....3P,2013ApJS..208....4P,2015ApJS..220...15P,2016ApJS..222....8D,2016ApJ...823..102C}, taking $v/v_{\rm crit}=0$ and [Fe/H]=0, which are in this case $R_*\approx\{.34,1.02,2.93\}R_\odot$. 
The differences between the relativistic and post-Newtonian treatments are clearest in the apsidal precession which can be underestimated by an order of magnitude as $r_p$ approaches $4r_g$. However, for such deep orbits precession does not occur spontaneously at periapse, as illustrated in Figure \ref{fig:orbit2}, so the post-Newtonian underestimation of $\Delta\phi$ is somewhat mitigated when the approximation of instantaneous precession is used to estimate $r_i$. The angle of the stream self-intersection however is vastly overestimated in the post-Newtonian approximation. Examining again Figure \ref{fig:orbit2}, it can be seen that as the degree of apsidal precession increases at high SMBH masses, the first stream self intersection becomes more grazing, although the subsequent self-intersections could occur at larger angles of intersection.

\begin{figure}
\includegraphics[width=\columnwidth]{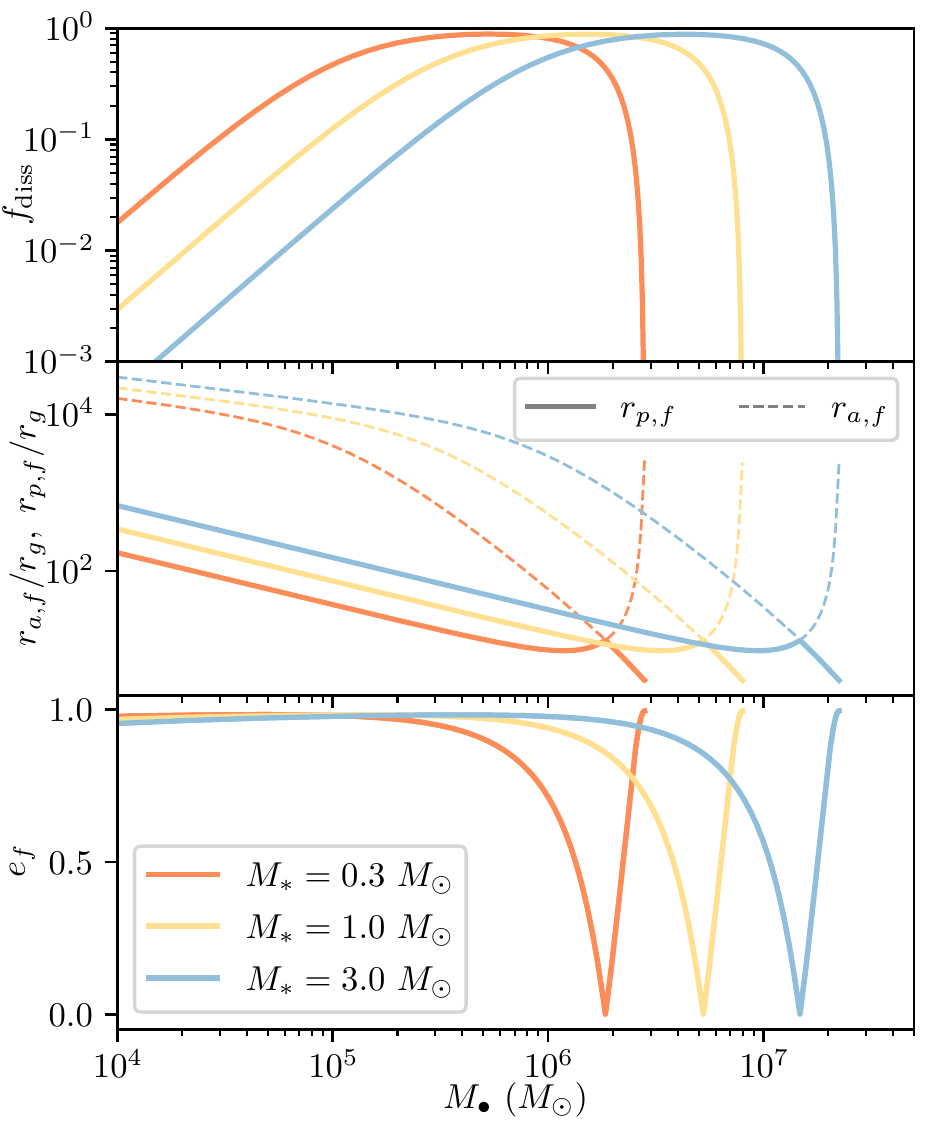}
\caption{The dissipated fraction of the orbital energy at the first stream self-intersection (top), post-intersection apoapse and periapse (middle), and post-intersection gas eccentricity (bottom) under the assumption of a completely inelastic collision. The stellar and orbital parameters are the same as in Figure \ref{fig:intersection}.}
\label{fig:PostIntersection}
\end{figure}

\section{Fully inelastic collisions}\label{sec:dissipation}
A common approximation of stream intersection dynamics is that the stream-stream collision is completely inelastic.
In the approximation that the incoming and outgoing streams at self-intersection have the same densities and cross sections, the streams then have equal azimuthal momenta and radial momenta that are equal in magnitude but opposite in sign. Then, the completely inelastic limit assumes the total cancellation of the gas radial velocity at constant angular velocity. 
The energy of the orbit post-intersection can be determined in a straightforward manner by evaluating Equation (\ref{eq:rdot2}) at the self-intersection radius: as the specific angular momentum is conserved and $\dot{r}=0$, the post-intersection specific energy can be simply calculated as $E_f^2=(1-r_s/r_i)(1+h^2/r_i^2)$. In analogy to \citep{2020MNRAS.492..686L}, we can define the fraction of dissipated energy relative to the orbital energy $f_{\rm diss}$ as 
\begin{equation}
f_{\rm diss}=\frac{E-E_f}{1-\sqrt{1-r_s/r_i}}.
\end{equation}

Given that the specific angular momentum of the stream is conserved during the collision, the trajectory of the material can be determined analytically by first solving for the roots of Equation (\ref{eq:dueq}). We denote the post-collision apoapse as $r_{a,f}$ and the post-collision periapse as $r_{p,f}$, defining the post-collision eccentricity as 
\begin{equation}\label{eq:ecc}
e_f = \frac{r_{a,f}-r_{p,f}}{r_{a_f}+r_{p,f}}.
\end{equation}
We remark that the simple geometric definition of eccentricity used here is, while still only a function of $E$ and $h$, more physically meaningful than a definition of eccentricity commonly used in the literature \citep[e.g.][]{2020MNRAS.492..686L,2021MNRAS.507.3207C}: $e=\sqrt{1+2(E-1)h^2r_g^{-2}}$. The latter expression can yield imaginary values even for geodesics with well-defined pericentres and apocentres. Take, as an example, the innermost stable circular orbit, which has $h=\sqrt{12}r_g$ and $E=\sqrt{8/9}$: in this case $1+2(E-1)h^2r_g^{-2}\approx-0.372$, whereas application of Equation (\ref{eq:ecc}) results in an eccentricity of zero, as expected for circular orbits. Analogously, we can calculate the semi-major axis of the post-intersection stream in a similarly general way: $a_f=(r_{a,f}+r_{p,f})/2$.

We display the post-intersection properties of the stream in Figure \ref{fig:PostIntersection} for three main sequence stars on $\beta=3$ orbits. Examining first $f_{\rm diss}$, we see that there are three regions which generally correspond to different regimes of apsidal precession. In cases where stars are disrupted many gravitational radii from the SMBH, self-intersection occurs near apoapse at mild angles, and relatively little energy is dissipated. As apsidal precession becomes greater, stream self-intersection occurs at smaller distances from the SMBH, and the radial component of the stream velocity is greater at self-intersection, leading to more significant dissipation. However, as the amount of apsidal precession further increases, the stream self-intersection again becomes more grazing and very little energy is dissipated. We emphasise however, that these small values of $f_{\rm diss}$ are only applicable to the \textit{first} stream self-intersection: as illustrated in Figure \ref{fig:orbit2}, multiple self-intersections can occur over the course of a single pericentre passage, and thus the reported values of $f_{\rm diss}$ may represent a lower limit. 

We present the post-intersection pericentric and apocentric distances together to illustrate visually their connection to the post-intersection gas eccentricity. Across most of parameter space, especially for orbits with pericentres $r_p\gg r_s$, the post-collision apocenter tracks the intersection radius and the post-collision pericentre tracks the initial pericentre distance. However, for orbits with pericentres nearer the SMBH, the pericentre must increase as the gas becomes more bound making the post-intersection gas much less eccentric. This can be understood from Equation (\ref{eq:h}): in the limit where $r_p\gg r_s,$ decreases in $E$ naturally lead to decreases in $r_p$ when $h$ is held constant as $h\propto r_p\sqrt{E^2 -1}$. However, when $r_p$ is comparable to $r_s$, $r_p$ must increase to hold $h$ constant while $E$ is decreased. 

\begin{figure}
\includegraphics[width=\columnwidth]{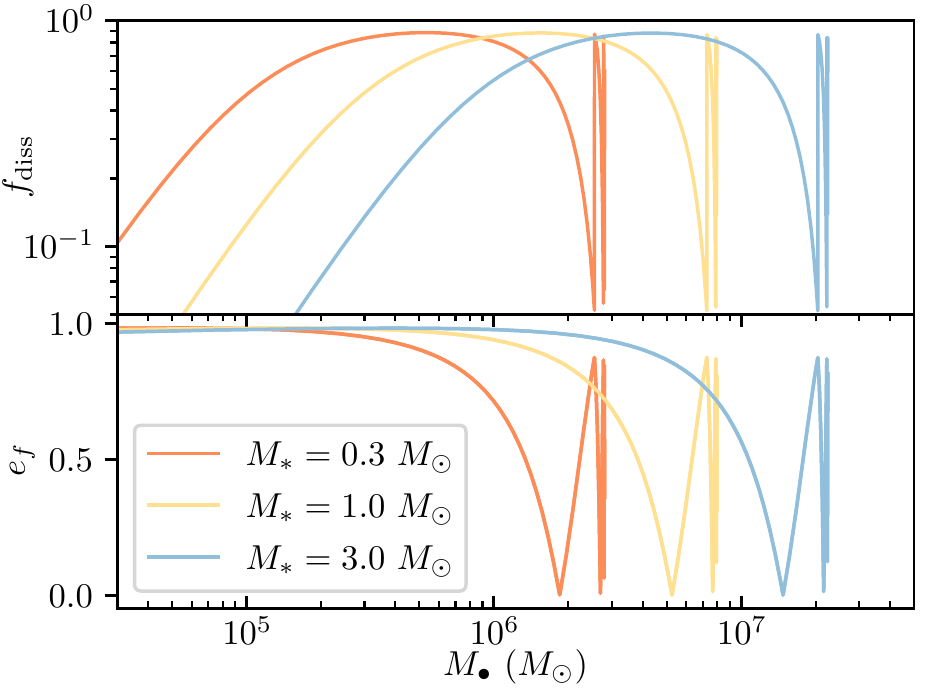}
\caption{The dissipated fraction of the orbital energy at the first stream pericentre passage (top), and post-intersection gas eccentricity (bottom) under the assumption of a completely inelastic collision, taking into account to most dissipative self-intersection if multiple occur in a single pericentre passage.  Note the different y-axis range in the top panel compared to Figure \ref{fig:PostIntersection}. The stellar and orbital parameters are the same as in Figure \ref{fig:intersection}.}
\label{fig:PostIntersectionMulti}
\end{figure}

\begin{figure}
\includegraphics[width=\columnwidth]{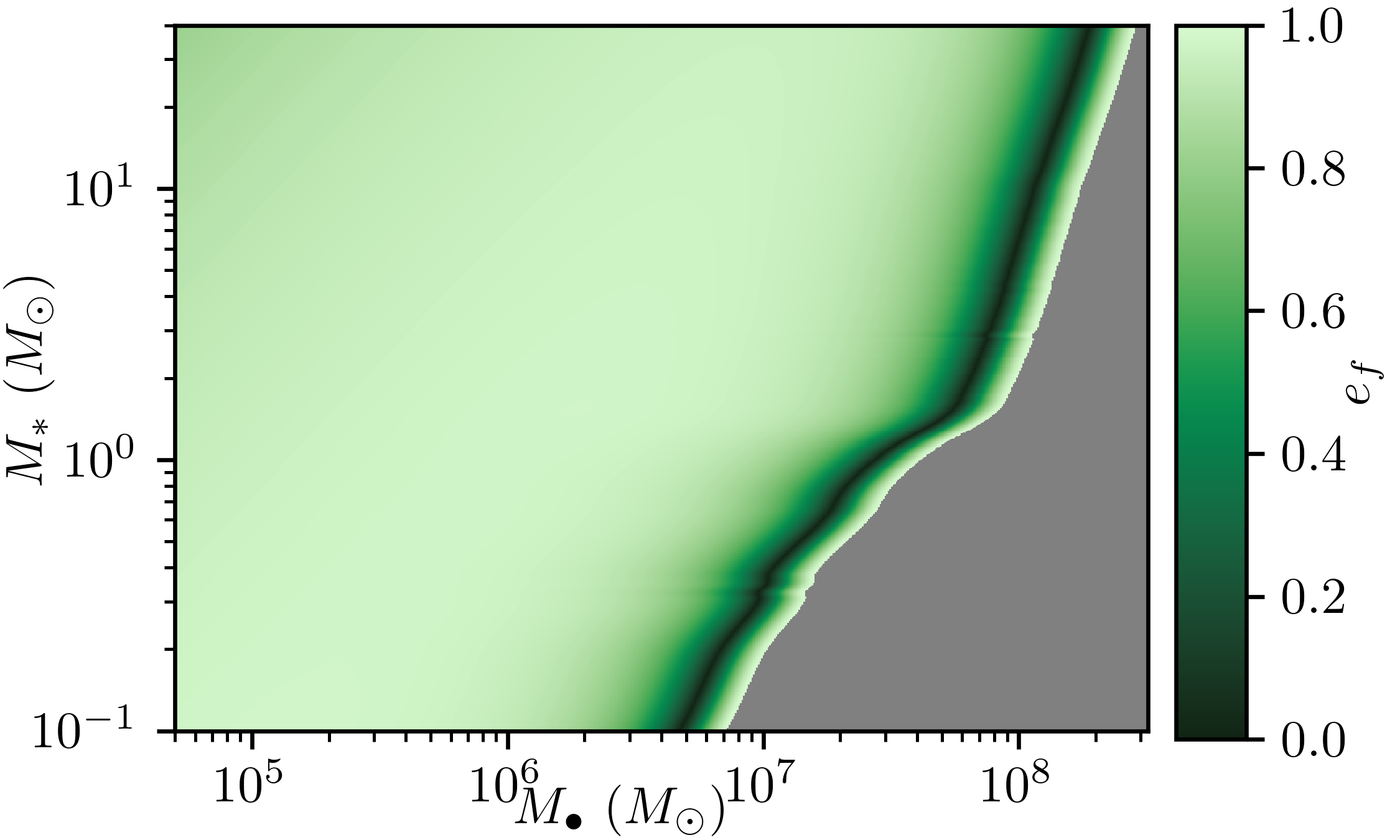}
\caption{The post-intersection eccentricity, considering only the first self-intersection, for the streams resulting from $\beta=1$ disruptions of intermediate age main sequence stars of various masses. The radii used here are derived from a 3rd-order spline interpolating the values from the MIST v1.2 evolutionary tracks \citep{2016ApJS..222....8D,2016ApJ...823..102C}. Regions where $r_p<4r_g$ are coloured grey.}
\label{fig:eGridMIST}
\end{figure}

\begin{figure}
\includegraphics[width=\columnwidth]{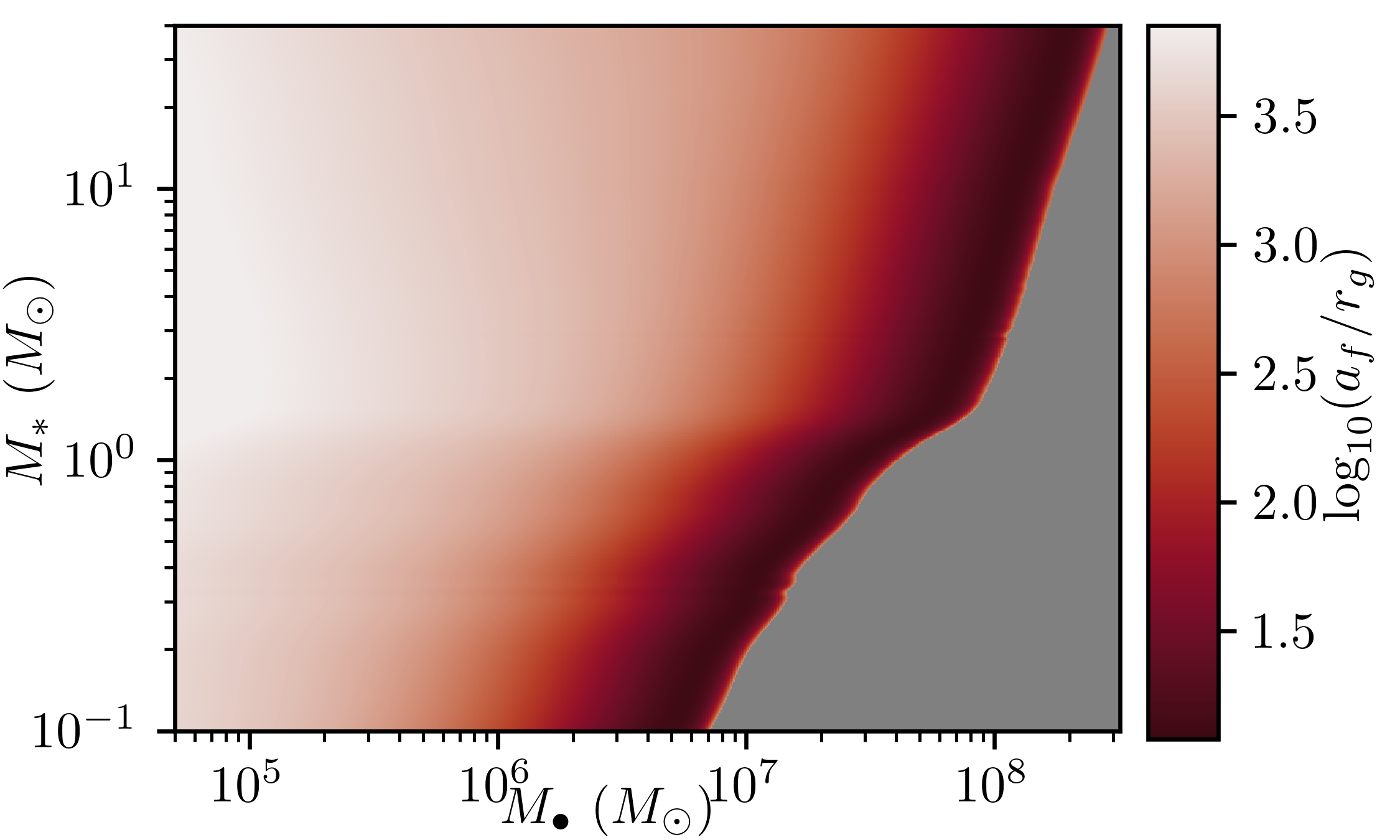}
\caption{The post-intersection stream semi-major axis, derived from the same inputs as Figure \ref{fig:eGridMIST}. Regions where $r_p<4r_g$ are coloured grey.}
\label{fig:aGridMIST}
\end{figure}

Evidently, at least under the assumption of an inelastic collision, there are a range of stellar and orbital parameters which lead not to the formation of a highly-eccentric disc, but instead a mildly-eccentric to circular disc after just a single pericentre passage of the stream. Figure \ref{fig:PostIntersection} shows the estimates of the post-intersection stream eccentricity $e_f$, but this is an upper limit on the eccentricity because of the possibility of multiple stream self-intersections during a single pericentre passage. We illustrate this in Figure \ref{fig:PostIntersectionMulti}, not by making calculations of multiple self-intersections, but simply by considering the most dissipative of the self-intersections over a single pericentre passage to calculate $f_{\rm diss}$ and $e_f$. Even under the assumption of only a single self-intersection each pericentre passage, it is apparent that a significant degree of dissipation takes place for streams which undergo strong precession. 

Because our methods are completely analytic, the calculation of large grids of TDE properties requires minimal effort. Accordingly, we plot the post-intersection stream eccentricities in Figure \ref{fig:eGridMIST} and semi-major axes in Figure \ref{fig:aGridMIST}, for main sequence stars between $0.1$ and $40~M_\odot$, assuming $\beta=1$ disruptions. These results as a function of stellar mass should be interpreted with some caution because of variation of the critical value of $\beta$ required for full disruptions, which would result in an effective weaker dependence on stellar mass \citep{2020ApJ...904...98R,2020ApJ...904..100R}. There is clearly a large parameter space for TDEs over which accretion discs can form promptly with semi-major axes $a_f\lesssim100~r_g$ and low-to-moderate eccentricities. The general trends in $e_f(M_\bullet)$ and $a_f(M_\bullet)$ with $M_*$ follow from $d\log{(R_*/R_\odot)}/d\log{(M_*/M_\odot)}$, which becomes notably greater below $\sim\!\!1\!-\!2~M_\odot$ due predominantly to the large convective envelopes in low-mass main sequence stars. However, it must be kept in mind that the results in this section were derived under the assumption of a completely inelastic collision, which is unlikely to hold in general. Furthermore,

\section{Relativistic shocks}\label{sec:shocks}
Although shocks during the self-intersection of TDE streams can both heat the gas and emit radiation which can potentially constitute a majority of the emission observed in optical and UV TDEs \citep[e.g.][]{2016ApJ...830..125J}, radiation hydrodynamic simulations suggest that only $\lesssim8\%$ of the kinetic energy of the stream is radiated, indicating largely elastic collisions \citep{2016ApJ...830..125J}. Such low radiative efficiencies can lead to a significant portion of the stream material becoming unbound, potentially covering a large solid angle and obscuring the smaller-scale accretion flow \citep{2016ApJ...830..125J,2020MNRAS.492..686L}. In the following we will consider the general properties of outflows from the self-intersection for the full range of shock efficiencies, considering together kinetic energy losses to both radiation and fluid internal energy. The calculations herein can be adapted in a straightforward manner to a given model for the distribution of outflowing material, which we illustrate using two example distributions. 

While the quantities determining the post-shock orbital characteristics of the stream are best calculated in a global frame, it is most straightforward to consider shock physics in the zero-net-momentum frame of the shock. Using the notation of \citet{2020MNRAS.492..686L}, the 4-velocity of the outward-moving stream in the frame of a stationary observer at the self-intersection radius is $\widetilde{u}_\alpha=(\widetilde{u}_t,\widetilde{u}_r,\widetilde{u}_\theta,\widetilde{u}_\phi)=\widetilde{\Gamma}(1,\widetilde{v}_r,0,\widetilde{v}_\phi)$, where $\widetilde{\Gamma}\equiv(1-\widetilde{v}_r^2-\widetilde{v}_\phi^2)^{-1/2}$, and $\widetilde{v}_r$ and $\widetilde{v}_\phi$ are given by Equations \ref{eq:vsquigr} and \ref{eq:vsquigp} respectively. Under the assumption that the radial momenta of the incoming at outgoing streams are equal in magnitude and opposite in sign, and that the angular momenta of those streams are equal, the zero-net-momentum (comoving) frame of the shock at $r_i$ moves with $\widetilde{v}_\phi$ in the $\phi-$direction relative to a stationary observer. In the comoving frame, in which we denote measurements of quantity $q$ by $\overline{q}$, the streams collide with radial 4-velocity components $|\overline{u}_r|=\widetilde{\Gamma}
~\!\!
\widetilde{v}_r,$ and collision velocity $\overline{v}_c=|\overline{u}_r|(1+\overline{u}_r^2)^{-1/2}.$

Local hydrodynamic simulations of stream self-intersection, both radiative and 3-dimensional \citep{2016ApJ...830..125J}, and non-radiative and 2-dimensional \citep{2020MNRAS.492..686L}, suggest that material flowing out from the self-intersection shock does so with a velocity distribution that is spatially uniform in the comoving frame. \citet{2016ApJ...830..125J} found that $\sim\!\!2$ to $\sim\!\!7$ per cent of the initial kinetic energy of the stream was lost as radiation, with an increasing radiated fraction at lower accretion rates, and found that although the internal thermal energy of the stream increased by orders of magnitude through shocking, the internal energy constituted a negligible portion of the total energy budget. With these results in mind, we assume that the magnitude of the velocity of outflowing material $\overline{v}_o$ is equal in all directions $(\overline{\theta},\overline{\phi})$ and given by $\overline{v}_o\equiv\alpha^{1/2}\overline{v}_c$ where $\alpha\in[0,1]$ is a dimensionless parameter controlling the dissipation (radiation) of energy during the shock. The completely inelastic collisions previously discussed are described by $\alpha=0$, while $\alpha=1$ corresponds to the completely elastic limit studied by \citet{2020MNRAS.492..686L}. 

The 4-velocity of the post-shock outflow in an arbitrary direction $\overline{\theta},\overline{\phi}$ in the comoving frame can be written in Cartesian components $\overline{u}_\alpha=\overline{\Gamma}(1,\overline{v}_x,\overline{v}_y,\overline{v}_z)$, where $\overline{\Gamma}=(1-\overline{v}_o^2)^{-1/2},$ $\overline{v}_x = \overline{v}_o\sin{\overline{\theta}}\cos{\overline{\phi}},$ $\overline{v}_y = \overline{v}_o\sin{\overline{\theta}}\sin{\overline{\phi}},$ and $\overline{v}_z = \overline{v}_o\cos{\overline{\theta}}.$ We orient our Cartesian coordinates such that the z-direction is perpendicular to the plane of the initial (prior to self-intersection) geodesic and that the azimuthal motion of both streams prior to the shock is in the x-direction, such that the y-direction corresponds to the direction towards or away from the black hole at the point of self intersection. Then, the corresponding 4-velocity components in the frame of a stationary observer at the intersection radius are $\widetilde{u}_\alpha(\overline{\theta},\overline{\phi})=(\widetilde{\Gamma}_\phi(\overline{u}_t+\widetilde{v}_\phi\overline{u}_x),\widetilde{\Gamma}_\phi(\widetilde{v}_\phi\overline{u}_t+\overline{u}_x),\overline{u}_y,\overline{u}_z),$ where $\widetilde{\Gamma}_\phi\equiv(1-\widetilde{v}_\phi^2)^{-1/2}$ is the Lorentz factor corresponding to the relative motion between the comoving and stationary frames. The post-shock specific energy ($E_p$) and specific angular momentum ($h_p$) in the global frame are then
\begin{equation}
E_p(\overline{\theta},\overline{\phi})=\mu_i^{1/2}\widetilde{u}_t
\end{equation}
and
\begin{equation}
h_p(\overline{\theta},\overline{\phi})=r_i\sqrt{\widetilde{v}_x^2+\widetilde{v}_z^2}.
\end{equation}

\begin{figure}
\includegraphics[width=\columnwidth]{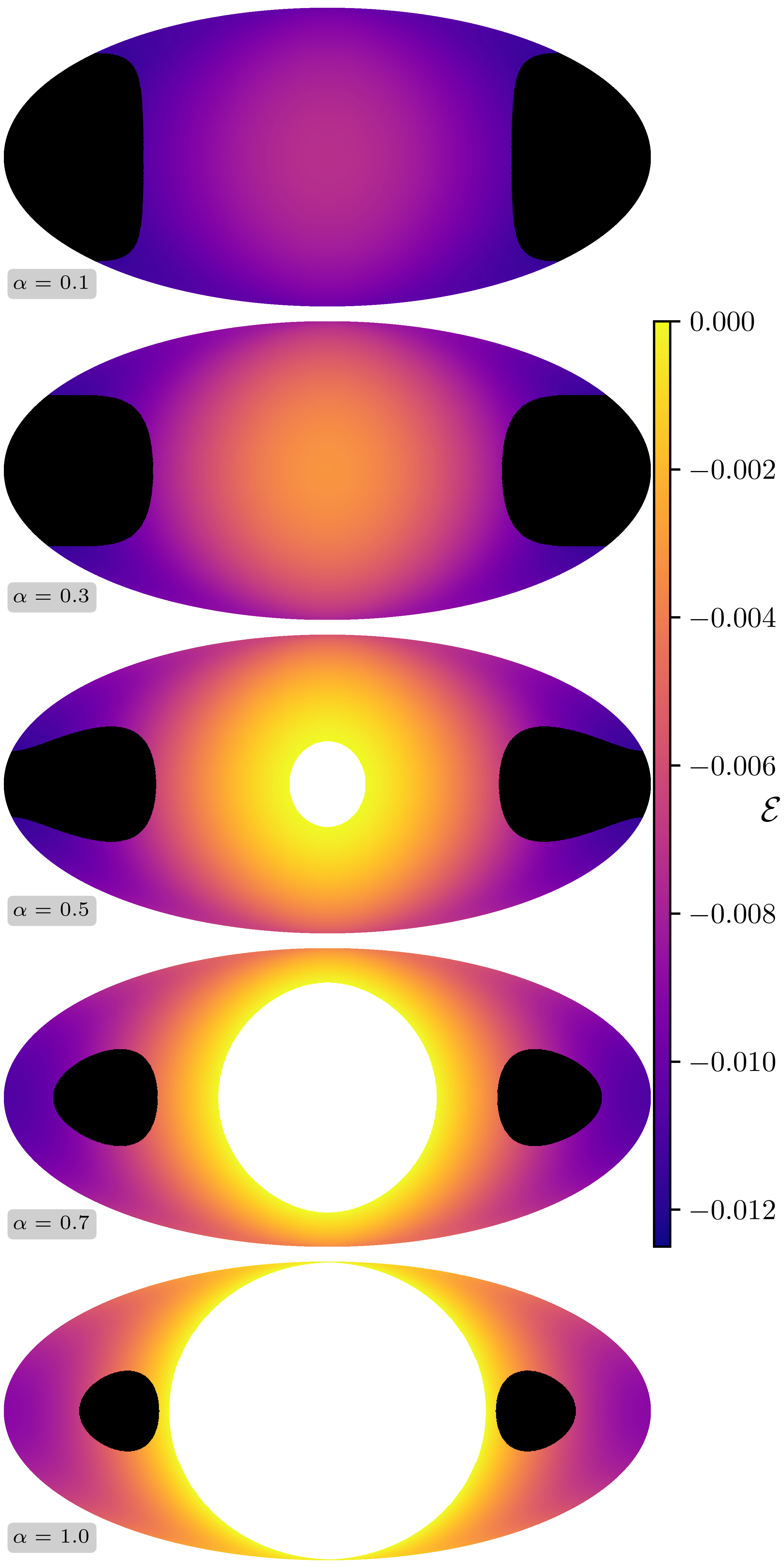}
\caption{The aftermath of the self-intersection shock of the stream resulting from the disruption of a $1~M_\odot$ star on a $\beta=3$ orbit by a $2.5\times10^6~M_\odot$ SMBH, shown via Mollweide projections of a spherical neighbourhood around the self-intersection point in the zero-momentum frame of the shock. Colours in shades of blue through yellow indicate the binding energy energy ($\mathcal{E}=E-1$) of the post-shock material in the global frame. White regions indicate unbound material, and material on plunging geodesics is shown in black. Each row plots results for a different collision efficiency $\alpha=\{0.1, 0.3, 0.5, 0.7, 1.0\}$ from top to bottom. Each Mollweide projection is oriented such that the equator lies the the plane of the initial orbit ($\overline{\theta}=\pi/2)$. The central meridian corresponds to $\overline{\phi}=0,$ and thus to a boost in the direction of the net angular momentum in the frame of a stationary observer. The left and right extrema ($\overline{\phi}=\pm\pi$) along the equator correspond to boosts in the direction opposite the net momentum in the stationary frame, while $\overline{\phi}=\pm\pi/2$ correspond to boosts in the radial direction.}
\label{fig:alphaMoll}
\end{figure}

Naturally, any material with $E_p>1$ ($\mathcal{E}>0$) is unbound from the black hole. To assess whether or not material is on a plunging geodesic, we evaluate the determinant of Equation (\ref{eq:dueq}) to determine the nature of its roots: cases where only one root ($u_3$) is real correspond to plunging worldlines, although subsequent collisions with disrupted material may interfere with such trajectories. 

We visualise in Figure \ref{fig:alphaMoll} the fate of post-shock gas following the disruption of a $1\,M_\odot$ star on a $\beta=3$ orbit by a $2.5\times10^6\,M_\odot$ SMBH, for a small set of shock efficiencies ($\alpha$). These plots are coloured based on the specific energy and angular momentum in the global Schwarzschild spacetime as functions of $\overline{\theta}$ and $\overline{\phi}$ in the frame of the collision. In these calculations we discretise both $\overline{\theta}$ and $\overline{\phi}$ using a few hundred increments. The equator corresponds to the plane of the initial geodesic of the disrupted TDE stream, and the central meridian corresponds to the direction of total momentum of the incoming and outgoing streams in the frame of a stationary observer at the self-intersection point. For higher values of $\alpha$, or shocks which convert small fractions of the stream kinetic energy into heat or radiation, material that is scattered in the direction of the net momentum of the stream can gain enough energy to become unbound. Material that acquires some momentum perpendicular to the net momentum of the intersecting streams, but which is still in the plane of the initial geodesic of the TDE stream, can be sent plunging into the SMBH due to the loss of angular momentum. On the other hand, material which acquires some out-of-plane velocity is less likely to plunge, instead orbiting in a separate plane than that of the pre-shock geodesic. 

\begin{figure}
\includegraphics[width=\columnwidth]{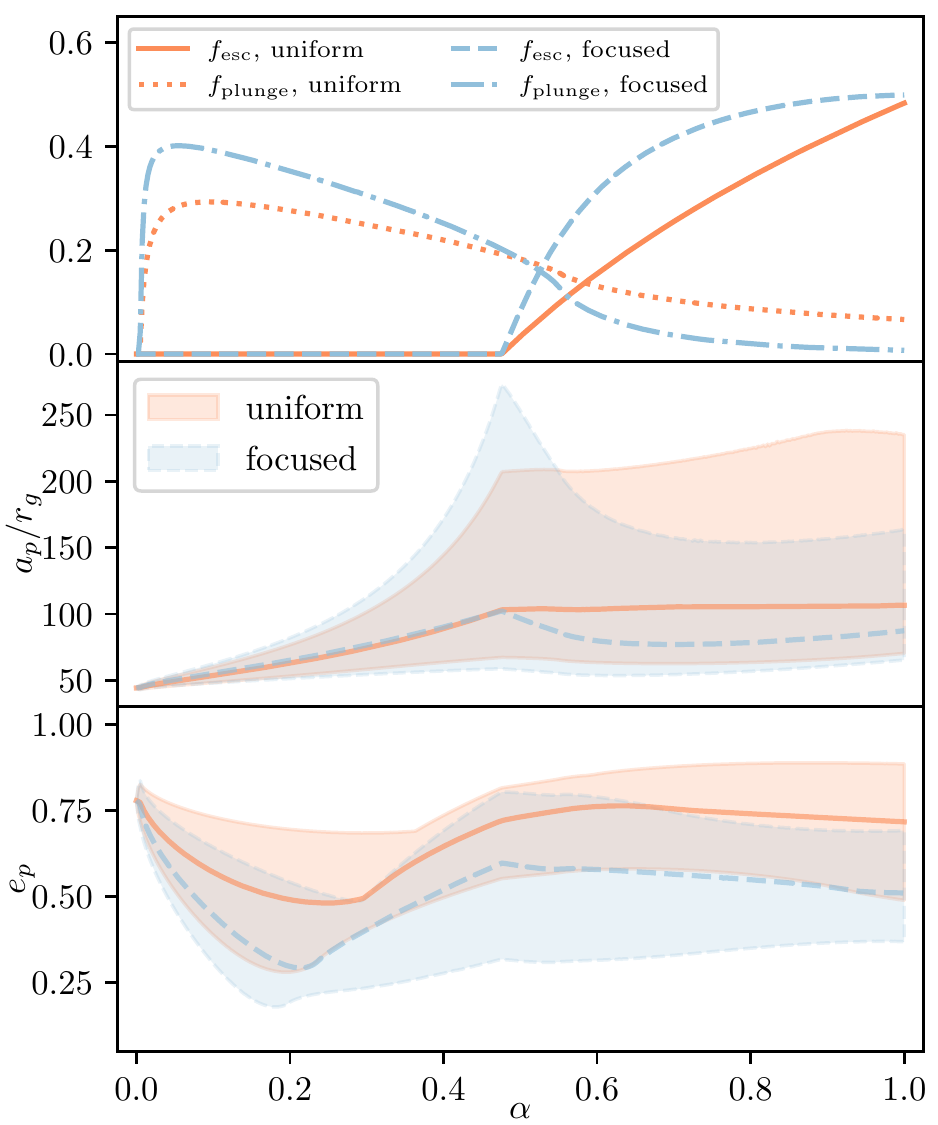}
\caption{Post-intersection stream properties as a function of $\alpha$ following the disruption of a $1~M_\odot$ star on a $\beta=3$ orbit by a $2.5\times10^6~M_\odot$ SMBH. The top panel shows the fractions of escaping (solid/dashed lines) and plunging (dotted/dash-dotted) material, the middle panel shows the characteristic semi-major axes of post-intersection debris, and the bottom panel shows the eccentricity distribution of post-intersection debris. In the second and third panels, the region between the 25th and 75th percentiles is shaded, and the lines denote the 50th percentile. Results assuming a uniform outflowing distribution of matter are shown in orange, and results assuming that more material outflows in directions perpendicular to the shock ($\propto\exp(-2\pi\sin^4{\overline{\phi}})$, `focused') are shown in blue.}
\label{fig:alphaAll}
\end{figure}

Once $E_p$ and $h_p$ are known for each direction, it is straightforward, at least as far as subsequent collisions can be neglected, to calculate the fraction of material which plunges into the black hole ($f_{\rm plunge}$) and that which escapes, becoming unbound ($f_{\rm esc}$). We calculate these quantities under two assumptions about the distribution of material after self-intersection: we consider material which moves away from the self intersection point with uniform density in $\overline{\theta},\overline{\phi}$ in the frame of the shock (`uniform'), and material which is preferentially ejected perpendicular to the direction of the pre-shock streams with an arbitrary weighting $\propto\exp(-2\pi\sin^4{\overline{\phi}})$ (`focused'). The plunging and escaping mass fractions are shown for both cases in Figure \ref{fig:alphaAll} for a $1\,M_\odot$ star on a $\beta=3$ orbit disrupted by a $2.5\times10^6\,M_\odot$ SMBH. For both mass distributions, material begins to be ejected around $\alpha\approx0.5$, and by $\alpha\approx1,$ $\sim\!50$ per cent of the material is ejected. Because $\alpha=0$ reduces to the completely inelastic case, both the escaping and plunging fractions must be $0$ or $1$. Because the TDE stream was bound initially, larger values of $\alpha,$ leading to larger scatter in velocities, are necessary for any material to escape. On the other hand, smaller values of $\alpha$ are able to remove enough angular momentum from material that it is sent plunging. 

Additionally, it is straightforward to use $E_p$ and $h_p$ to calculate the post-shock eccentricity semi-major axis of geodesics, $e_p(\overline{\theta},\overline{\phi})$ and $a_p(\overline{\theta},\overline{\phi})$ along each discrete post-shock trajectory. Using these, we calculate characteristic post-shock eccentricities and semi-major axes of the material which is accreted, i.e. the bound material which is not on a plunging geodesic. In doing so, we can estimate the the properties of the accretion disc formed after the self-intersection shock. Because there is a distribution of $e_p$, $a_p$ values for $\alpha>0$, we calculate the 25th, 50th, and 75th percentiles of these quantities, which are presented in Figure \ref{fig:alphaAll}. Turning first to the characteristic semi-major axes of post-shock geodesics, we see that for $0<\alpha\lesssim0.5$ semi-major axes tend to increase with increasing $\alpha$. However, because these probability distributions include only the material that does not promptly plunge or become ejected from the system, once $f_{\rm esc}$ becomes greater than zero, the average debris semi-major axis becomes roughly constant, or declines for some intermediate values of $\alpha$. Turning to the post-shock eccentricity, regardless of the value of $\alpha$ or the post-shock material profile, the average eccentricity of \textit{accreting} material tends to be lower than the value corresponding to a completely inelastic collision ($\alpha=0$) for this SMBH mass. On the other hand, in cases where a completely inelastic collision would result in $e_p\sim0$ orbits, more elastic collisions result in higher-eccentricity accreting debris. Post-shock material with a `focused' density distribution tends to have slightly lower eccentricities than material with a uniform angular distribution in the shock frame.

\begin{figure}
\includegraphics[width=\columnwidth]{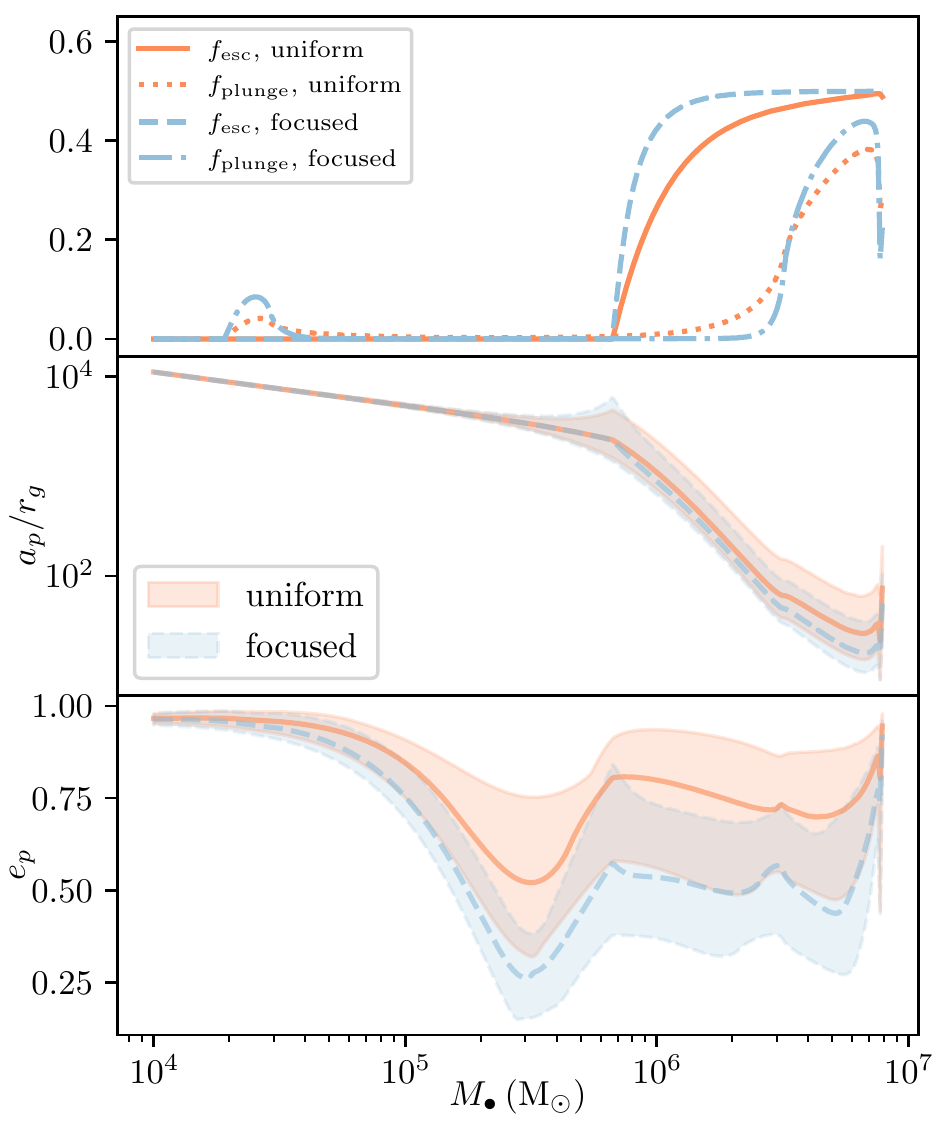}
\caption{The fractions of escaping (solid/dashed lines) and plunging (dotted/dash-dotted) material (top panel), characteristic semi-major axis (middle panel), and eccentricity (bottom panel) of accreting (neither plunging nor unbound) material as a function of SMBH mass ($M_\bullet$) after the self-intersection of the stream resulting from the disruption of a $1~M_\odot$ star on a $\beta=3$, assuming a shock efficiency $\alpha=0.95$. Results assuming a uniform outflowing distribution of matter are shown in orange with solid lines, and results assuming that more material outflows in directions perpendicular to the shock ($\propto\exp(-2\pi\sin^4{\overline{\phi}})$, `focused') are shown in blue with dashed lines. The shaded envelopes indicate the 25th to 75th percentiles, and the inner lines indicate the 50th percentile.}
\label{fig:shockSMBH}
\end{figure}

Although lower values of $\alpha$ may be associated with low accretion rates, either during the later stages of a full disruption or following a partial disruption, values of $\alpha \gtrsim 0.9$ are likely to be the most astrophysically relevant \citep[e.g.][]{2016ApJ...830..125J}. 
Accordingly, we investigate post-shock characteristic geodesic semi-major axes and eccentricities over a range of SMBH masses in Figure \ref{fig:shockSMBH}, along with the fractions of unbound and plunging material, considering a solar-mass star on a $\beta=3$ orbit and an $\alpha=0.95$ self-intersection shock. Both the fraction of plunging and escaping material depend strongly on the SMBH mass though its influence on the self-intersection radius. We observe the same trends as \citet{2020MNRAS.492..686L}, such as the small bump in the plunging fraction of material at low masses, where $\widetilde{\theta}\sim\pi/4$ and thus the radial velocity and azimuthal velocity are quite similar, the sharp increase in the fraction of unbound material for SMBH masses as $r_i/r_g$ becomes smaller and the velocities of the colliding streams become larger, and the increased plunging fraction at higher masses still. The dependence on the self-intersection radius of the escaping fraction can help explain why some TDE simulations have identified outflows following the self-intersection shock \citep[e.g.][]{2016MNRAS.458.4250S} while other have not \citep[e.g.][]{2015ApJ...804...85S}.

Turning to the average properties of the accreting material, that which neither escapes nor directly plunges, we find a significant range of SMBH masses over which the resulting accretion disc has a low-to-moderate eccentricity, dropping to a median eccentricity of $\sim0.25$ in the `focused' case around $M_\bullet \sim 3\times 10^5\,{\rm M_\odot}$. Similarly to the $\alpha=0$ case shown in Figure \ref{fig:PostIntersection}, self-intersections very near the SMBH which occur at grazing angles due to the large precession for $M_\bullet \gtrsim 7 \times 10^6\,{\rm M_\odot}$ suggest that post-intersection eccentricities may remain high ($e_p\gtrsim0.6$) and the fraction of plunging material may be reduced, although the possibility of multiple self-intersections during a single pericentre passage makes such high eccentricity post-intersection accretion discs unlikely. 

\begin{figure}
\includegraphics[width=\columnwidth]{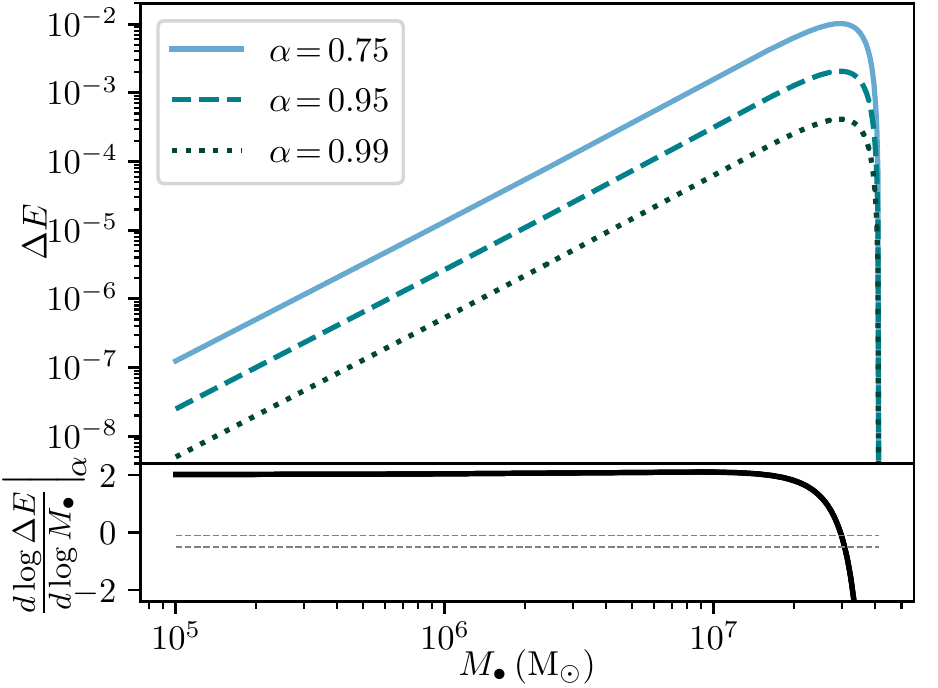}
\caption{Top panel: the specific energy dissipated ($\Delta E$), as measured in the global Schwarzschild spacetime, by the self-intersection shock of a $1\,M_\odot$ star on a $\beta=1$ orbit as a function of SMBH mass. The blue solid line plots results for $\alpha=0.75$, the medium green dashed line plots results assuming $\alpha=0.95$, and the dark green dotted line plots results assuming $\alpha=0.99$. Bottom panel: the power law exponent of $\Delta E(M_\bullet)$, $d\log{\Delta E}/d\log{M_\bullet}$ at constant $\alpha$ is shown by the black slolid line. Grey dashed lines indicate the range of values consistent with \citet{2020ApJ...894L..10H} (see Section \ref{sec:obs}).}
\label{fig:shockDE}
\end{figure}

In addition to typical post-shock gas characteristics, the energy dissipated during the self-intersection shock is also of astrophysical interest, as it may be responsible for some of the observed optical emission from many TDEs \citep[e.g.][]{2015ApJ...806..164P,2016ApJ...830..125J,2017MNRAS.467.1426S}. We show this radiated specific energy ($\Delta E$) for various SMBH masses and values of $\alpha$ in Figure \ref{fig:shockDE}, along with the power law index $d\log{\Delta E}/d\log{M_\bullet}$ at constant $\alpha$. At low masses the change in specific energy due to the self-intersection shock scales approximately as $\Delta E\propto M_\bullet^2$, which follows from 1st-order post-Newtonian estimates of the collision \citep[e.g.][]{2015ApJ...812L..39D,2020MNRAS.492..686L}, but for highly relativistic orbits far less energy is dissipated due to the grazing nature of the self-intersection. However, the latter scenario corresponds to orbits which undergo extreme apsidal precession such as that shown in Figure \ref{fig:orbit2}, and such streams would self-intersect multiple times if uninterrupted. Thus, in analogy to the results shown in Figure \ref{fig:PostIntersectionMulti}, the dissipated energy reported in Figure \ref{fig:shockDE} may be considered an estimate of a lower limit.

\begin{figure*}
\includegraphics{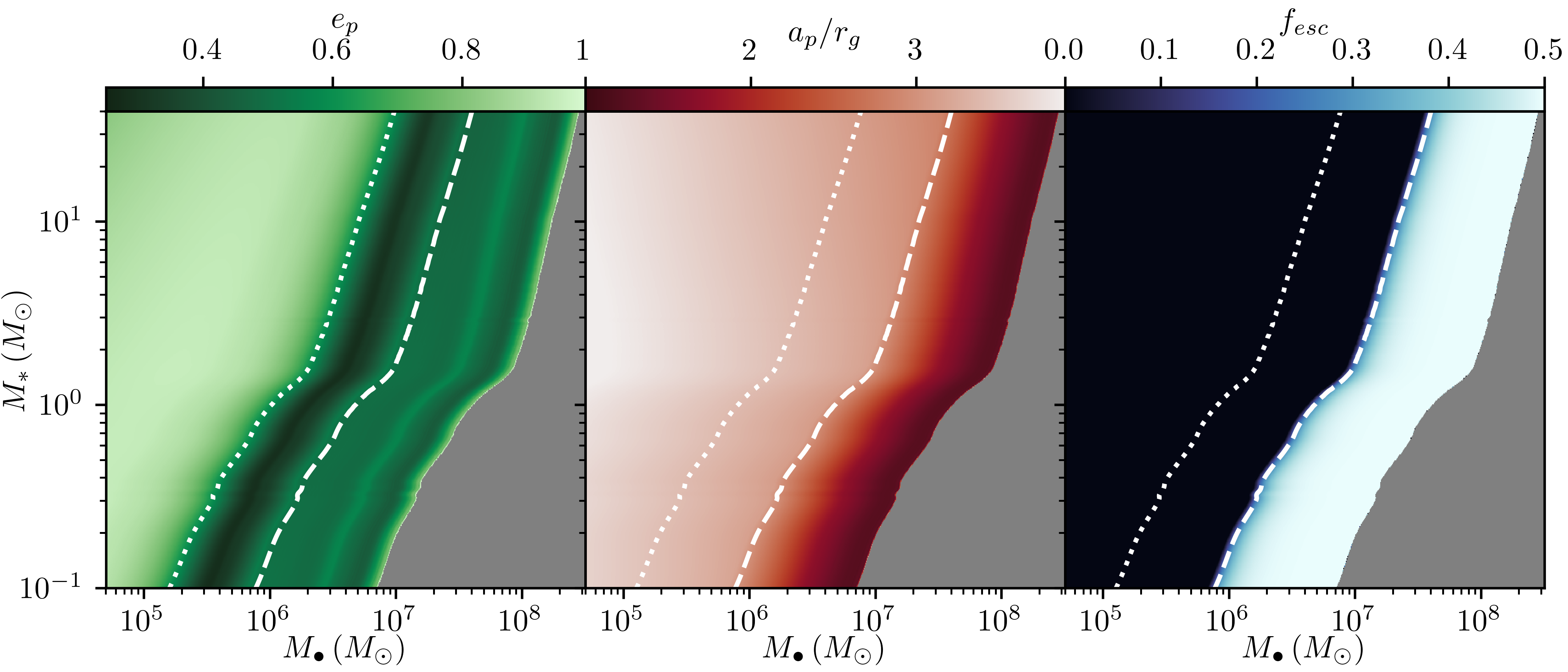}
\caption{The median post-intersection eccentricity of accreting material (left), semi-major axis of the accreting material (centre), and fraction of escaping material (right), following $\beta=1$ disruptions of main sequence stars. The dotted lines indicate the low-$M_\bullet$ region where $e_p=0.6$, an arbitrary choice made to differentiate regions where $e_p\sim1$ from where it is smaller. The dashed lines demarcate regions where $f_{\rm esc}$ is greater than or less than $0.25$. We have neglected to include a dotted line for the high-eccentricity post-intersection debris at very high SMBH masses for which the self-intersection collision is largely grazing.}
\label{fig:shockGrids}
\end{figure*}

We now turn to self-intersection shock outcomes as functions of both stellar mass ($M_*$) and SMBH mass ($M_\bullet$). For these calculations, we assume a `focused' distribution of outflowing matter in the shock frame ($\rho\propto\exp(-2\pi\sin^4{\overline{\phi}})$) where matter is preferentially directed perpendicularly to the incoming streams, which is qualitatively inspired by the stream self-intersection simulations presented in \citet{2016ApJ...830..125J} and \citet{2020MNRAS.492..686L}, but is certainly not in precise agreement with physical reality. Additionally, inspired by the range of radiative efficiencies observed in the simulations of \citet{2016ApJ...830..125J}, we fix $\alpha=0.95$ but note that for the `focused' mass distribution, $f_{\rm esc}$, $a_p$, and $e_p$ do not change significantly over the range $0.9<\alpha<1$. Using the intermediate age main sequence radii from the MIST evolutionary tracks \citep{2016ApJS..222....8D,2016ApJ...823..102C} as before, we show the median disc eccentricity, median semi-major axis, and fraction of escaping material in Figure \ref{fig:shockGrids}. We caution that we have held $\beta=1$ in Figure \ref{fig:shockGrids} which is not representative of realistic full disruptions \citep{2020ApJ...904...98R,2020ApJ...904..100R}.

The post-intersection disc has fairly mild eccentricity $e_p < 0.6$ over a large region of the $M_*-M_\bullet$ parameter space, dropping as low as $\sim\! 0.3$ for some masses. Part of this region coincides with the ranges of masses where almost no material is launched as an outflow, potentially leading to low-eccentricity accretion flows with little-to-no obscuring material. However, these discs would still have fairly large semi-major axes, $a_p \sim 10^3 r_g$. For disruptions by higher-mass SMBHs at fixed stellar mass, characteristic post-intersection disc semi-major axes can range from $10-100\,r_g$, although these cases typically coincide with orbital parameters for which roughly half of the material becomes unbound during the self-intersection shock.

\section{Discussion}\label{sec:discuss}
The eccentricity, semi-major axis, and ejected fraction of post-shock stream debris have a number of implications for the observable properties of TDEs and the nature of the corresponding accretion flows. Herein we discuss some of these implications, as well as some limitations of our present study. 

\subsection{Observational Trends}\label{sec:obs}
Thanks to  large surveys such as ASAS-SN \citep{2017PASP..129j4502K}, ZTF \citep{2019PASP..131a8002B}, and the all-sky X-ray survey carried out by the eROSITA telescope \citep{2014MNRAS.437..327K,2021A&A...647A...1P}, the number of tidal disruption events observed continues to grow, and trends in the population are beginning to emerge. 

For example, \citet{2020ApJ...894L..10H} found that in a sample of UV/optical TDEs, events with greater peak luminosities tend to dim over longer timescales. Similarly, \citet{2021ApJ...908....4V} found in their analysis of ZTF data that events with larger peak luminosities tend to have longer rise timescales leading up to their peak. Quantitatively, \citet{2020ApJ...894L..10H} found that the peak luminosity was related to the supermassive black hole mass by $L_{\rm peak}\propto M_\bullet^{-0.8\pm0.2}$, under the assumption that the decline timescale in the light curve follows the fallback timescale $\propto M_\bullet^{1/2}$. 

If the optical light curves of TDEs are dominated by stream self-intersection, only a narrow range of SMBH masses is in agreement with observed trends. In this case, the self-intersection luminosity can be calculated along the lines of $L_{\rm se}\propto (1-\alpha) \Delta E \dot{M} \propto (1-\alpha) \Delta E M_\bullet^{-1/2} M_*^2R_*^{-3/2}$. Thus, for a given stellar mass and shock efficiency,\footnote{A decrease in $\alpha$ (increased fraction of energy radiated by the collision) with decreasing accretion rate \citep[e.g.][]{2016ApJ...830..125J} would require more steep slopes for constant stellar properties.} the trends observed by \citet{2020ApJ...894L..10H} can be explained by stream self-intersections in the regime where $\Delta E \propto M_\bullet^{-0.3\pm0.2}$.
We compare the theoretical power-law index $d\log{\Delta E}/d\log{M_\bullet}$  to the values inferred from \citet{2020ApJ...894L..10H} in Figure \ref{fig:shockDE}. 
Due to the narrow range of SMBH masses for a given $M_*$ and $\beta$ where $\Delta E$ has a slope compatible with observations, fine tuning seems to be required for stream self-intersection shocks to be the dominant driver of TDE optical luminosities, although more general processes which track the accretion rate $(L\propto M_\bullet^{-1/2})$ seem in adequate agreement. 

Only a small fraction of TDEs identified via optical variability have also been visible in the X-rays \citep[e.g.][]{2021ApJ...908....4V}, which has been attributed to the obscuration and reprocessing of X-rays by an outflowing obscuring medium \citep[e.g.][]{2018ApJ...859L..20D,2020MNRAS.492..686L}. However, the all-sky X-ray survey carried out by 
SGR/eROSITA has recently discovered a sample of thirteen X-ray tidal disruption events, only four of which showed any optical activity \citep{2021MNRAS.508.3820S}. We remark that there is a large parameter space available for TDEs which produce no unbound material during stream self-intersection, shown to the left of the dashed lines in Figure \ref{fig:shockGrids}. For disruptions so far from the relativistic regime e.g. $M_\bullet \lesssim 4\times10^6\,M_\odot$ for $1\,M_\odot$ stars on $\beta=1$ orbits, typical self-intersection luminosities would be reduced by $\gtrsim 3$ orders of magnitude compared to more relativistic self-intersection shocks (c.f. Figures \ref{fig:shockDE} and \ref{fig:shockGrids}). Additionally, a sizeable fraction of the parameter space for self-intersection shocks which result in no unbound material also results in relatively low-eccentricity debris, as illustrated by the area between the dashed and dotted lines in Figure \ref{fig:shockDE}, which could potentially limit the optical emitted light by disc circularisation.

\subsection{Accretion physics}
The properties of accretion discs with nonzero eccentricity differ significantly from those of circular discs. For example, the phenomenology of the of the magnetorotational instability \citep{velikhov59,1960PNAS...46..253C,1998RvMP...70....1B} changes significantly in eccentric discs, where both the number of unstable modes, the wavenumber of the most unstable mode, and angular momentum flux as a function of orbital phase for the most unstable mode vary significantly as a function of disc eccentricity \citep{2018ApJ...856...12C}. Furthermore, eccentric accretion discs experience vertical oscillations which can become supersonic and lead to significant shocking, especially as disc eccentricity increases \citep{2021ApJ...920..130R}. 

If eccentric discs are common, it will be important to carry out representative simulations. For example, \citet{2020MNRAS.495.1374B} used smoothed particle hydrodynamics simulations to study the formation and evolution of eccentricity discs by injecting matter from the self-intersection shock as modelled by \citet{2020MNRAS.492..686L}, finding that after roughly a third of the orbital period of the most bound debris average particle eccentricities were about $\sim\!\!0.2$, despite initially being peaked around $e\sim\!\!1$. \citet{2021MNRAS.507.3207C} took a different approach, injecting streams of disrupted material and following the stream through both self-intersection and subsequent circularisation, using general relativistic grid-based simulations of radiation magnetohydrodynamics. Some earlier work \citep[e.g.][]{2018ApJ...859L..20D} assumed rapid circularisation and directly carried out general relativistic radiation magnetohydrodynamic simulations of compact super-Eddington accretion flows. Given that low-eccentricity accretion discs can form promptly after self-intersection even before further circularisation, such treatments may be appropriate over a wide range of stellar and SMBH parameters as shown in Figure \ref{fig:shockGrids}. However, for higher-mass stars at a given SMBH mass the post-intersection accretion flows may still have very high eccentricities, necessitating the study of a wide range of disc morphologies to understand the full population of TDEs.

\subsection{Limitations}
One of the foremost shortcomings of this study is our assumption of the Schwarzschild spacetime, while in general SMBHs are thought to have nonzero spin. We have made this assumption primarily for simplicity: \citet{2009CQGra..26m5002F} have derived closed-form solutions for timelike geodesics in Kerr spacetime. However, in application to tidal disruption events, nonzero spin not only adds the additional parameter of the dimensionless SMBH spin $(a_\bullet)$, but also the finite width of the TDE stream and the relative inclination between the stellar orbital angular momentum and SMBH spin ($i$). Thus, extension of our analytic framework to spinning black holes is well within the realm of reason and possibility for future studies, but beyond the scope of this work. 

One concern for TDE streams orbiting SMBHs with nonzero spin is that of Lens-Thirring precession, which arises for orbits the angular momentum of which is misaligned with the SMBH spin, which can delay stream self-intersection \citep[e.g.][]{1994ApJ...422..508K,2013ApJ...775L...9D,2015ApJ...809..166G,2016MNRAS.461.3760H,2021arXiv211203918B}. To first post-Newtonian order, the nodal Lens-Thirring precession over the course of one orbit is $\Delta \theta_{\rm LT}\approx 4\pi a_\bullet(r_s/4r_p)^{3/2}\sin{(i)}$. For $a_\bullet \ll 1$, or for $r_p \gg r_s$, Lens-Thirring precession is negligible and streams can intersect promptly for most stream thicknesses ($H/r\sim r_*/r_p$). On the other hand, \citet{2015ApJ...809..166G} found that for $a_\bullet \gtrsim 0.2$ self-intersection of the most-bound debris could be delayed by tens of orbital periods. However, because the degree of apsidal precession is typically much larger than that of Lens-Thirring precession, the actual dynamics of the stream self-intersection are largely unchanged once it finally occurs, apart from the differing intersection cross section \citep[e.g.][]{2015ApJ...809..166G,2020MNRAS.492..686L}. Additionally, spinning black holes can have smaller, or larger, marginally bound radii depending on the orientation of the SMBH spin ($-1\leq a_\bullet\leq 1$). In the case where the SMBH spin is parallel to the plane of the stellar orbit, the most bound radius is given by $r_{\rm mb}=(1+\sqrt{1-a_\bullet})^2r_g$ \citep{1972ApJ...178..347B}, which can facilitate disruptions around more massive SMBHs, altering disruption demographics.

Our calculations have also assumed that the cross section, binding energy, density, and radial velocity magnitude of the incoming and outgoing streams at the self-intersection radius are equal. The assumptions that the density, radial velocity magnitude, and specific energy of the incoming and outgoing streams are reasonable when self-intersection happens near pericentre, since the time for pericentre passage is orders of magnitude faster than the orbital timescale for high-eccentricity orbits. On the other hand, these approximations may be less reasonable in cases where self-intersection occurs near apocentre. It is also possible that viscous, magnetohydrodynamic \citep{2018ApJ...856...12C}, or hydrodynamic effects such as shocks \citep{2021ApJ...920..130R} which occur near pericentre could potentially lead to differences in the cross sections of incoming and outgoing streams. The extent to which these effects alter gas dynamics during the first pericentre passage of the disrupted material should be the subject of future studies. 

\section{Conclusions}\label{sec:conclusions}
We have developed an analytic, fully relativistic framework for studying the self-intersection of TDE streams. This framework, based on the analytic solution for Schwarzschild geodesics \citep{1930JaJAG...8...67H}, and building on previous post-Newtonian \citep{2015ApJ...812L..39D} and numerical works \citep{2020MNRAS.492..686L}, provides closed-form and exact expressions for the TDE stream self-intersection radius, geodesic semi-major axis, and eccentricity, among others. We have demonstrated this framework by calculating the properties of TDE streams at their self-intersection, as well as in predicting the properties of the accretion flow, and potentially outflow, following stream self-intersection. 

Based on comparisons between predictions from our self-intersection model with observations, we find that weakly-relativistic TDEs are able to produce events with comparatively small luminosity due to self-intersection, and no material unbound by the self-intersection shock, possibly explaining the large population of events identified by SGR/eROSITA unaccompanied by counterparts. With application to both optical luminosity stemming from disc circularisation and physics of the potentially eccentric discs formed after stream self-intersection, we have identified regimes where characteristic post-intersection disc eccentricities can reach $\sim 0.3-0.6,$ leaving less energy available for further circularisation at constant specific angular momentum, and changing the phenomenology of the MRI in the post-intersection disc. We have also found that trends in the peak luminosity and decay timescale of optical TDE light curves can be explained by the production of optical emission during the self-intersection shock for a narrow range of SMBH and stellar masses, but the fine tuning required to reproduce the strength of the observed correlation may disfavour self-intersection shocks as the dominant origin of optical TDE emission. 

\section*{Acknowledgements}
This work was inspired by simulations of TDE streams performed by Fraol Kebede as part of the 2021 GRAD-MAP Summer Scholars Program at the University of Maryland.\footnote{\url{https://www.umdgradmap.org/}}
I am grateful for stimulating discussions on TDEs with Brad Cenko, Erica Hammerstein, Cole Miller, Brenna Mockler, and Geoff Ryan. I am also thankful for feedback on an earlier version of this manuscript from Brad Cenko, Jane Dai, Julian Krolik, and Cole Miller.
This work was supported by NASA ADAP grant 80NSSC21K0649.

\section*{Data Availability}
The data underlying this article will be shared upon reasonable request to the author.

\bibliographystyle{mnras}
\bibliography{references.bib}


\appendix
\section{Orbit equation derivation}\label{app:deriv}
As the solution to Equation (\ref{eq:dueq}) may not be obvious, we provide a brief derivation of Equation (\ref{eq:schworbit}) under the simplifying assumption that all three roots are real. We can write Equation (\ref{eq:dueq}) as 
\begin{equation}\label{eq:cubic}
\mathcal{P}=\left(\frac{du}{d\phi}\right)^2=a_3u^3+a_2u^2+a_1u+a_0,
\end{equation}
where $a_3=r_s$, $a_2=-1$, $a_1 = r_s/h^2$, and $a_0=(E^2-1)/h^2$, or in terms of its roots
\begin{equation}\label{eq:factCubic}
\mathcal{P}=A(u-u_1)(u-u_2)(u-u_3),
\end{equation}
where $A=a_3=r_s$. To find explicit forms of the roots, we can rewrite Equation (\ref{eq:cubic}) in its depressed form, defining $g\equiv u+a_2/(3a_3)$, as
\begin{align}\label{eq:depressCubic}
\mathcal{P}=a_3(g^3+pg+q),\\p=\frac{3a_1a_3-a_2^2}{3a_3^2},\\q=\frac{2a_2^3-9a_1a_2a_3+27a_0a_3^2}{27a_3^3}.
\end{align}
All three roots are real when $p<0,$ in which case the roots are given by
\begin{equation}
u_k = -\frac{a_2}{3a_3} + 2\sqrt{\frac{-p}{3}}\cos{\left[\frac{1}{3}\arccos{\left(\frac{3q}{2p}\sqrt{\frac{-3}{p}}\right)}+\frac{2\pi k}{3}\right]},
\end{equation}
for $k\in\{1,2,3\}$ \citep{vietefrancisci}, which leads to $u_1\leq u_2 \leq u_3$ as desired.

Now that we have closed-form, albeit cumbersome, expressions for the roots of $\mathcal{P}$, we can solve
\begin{equation}
\frac{du}{\sqrt{(u-u_1)(u-u_2)(u-u_3)}}=\sqrt{A}d\phi,
\end{equation}
by first changing variables using 
\begin{align}
u=u_1+(u_2-u_1)\sin^2{\lambda}\\
du=2(u_2-u_1)\sin{\lambda}\cos{\lambda}d\lambda,
\end{align}
where $\lambda$ increases monotonically with $\phi$ such that
\begin{equation}\label{eq:factored}
\frac{2d\lambda}{1-k^2\sin^2{\lambda}}=\sqrt{A(u_2-u_1)}d\phi,
\end{equation}
where we have defined $k\equiv\sqrt{(u_2-u_1)(u_3-u_1)}$ as in Section \ref{sec:geodesics}. Because $0<k<1$, Equation (\ref{eq:factored}) can be integrated, resulting in 
\begin{equation}
\mathcal{F}(\lambda,k)=\int_0^\lambda\frac{d\zeta}{\sqrt{1-k^2\sin^2{\zeta}}}=\frac{1}{2}\phi\sqrt{A(u_3-u_1)},
\end{equation}
where $\mathcal{F}(\lambda,k)$ is the incomplete elliptic integral of the first kind and we have set the integration constant to zero. Then, as the Jacobi elliptic functions are inverses of elliptic integrals, we have 
${\rm sn}\left(\mathcal{F}(\lambda,k),k\right)=\sin{\lambda}$, and thus
\begin{equation}
u(\phi)=u_1+(u_2-u_1){\rm sn}^2\left(\frac{1}{2}\phi\sqrt{r_s(u_3-u_1)},k\right).
\end{equation}


\bsp	
\label{lastpage}
\end{document}